\documentstyle[12pt]{article}

\def\bra#1{\langle#1\vert}        %% bra vector  <w|
\def\ket#1{\vert#1\rangle}        %% ket vector |z>
\def\be{\begin{equation}}
\def\ee{\end{equation}}
\def\bea{\begin{eqnarray}}
\def\eea{\end{eqnarray}}%%%%
\def\hL{\hat{L}}                    %%operator L
\def\hw{\hat{w}}                    %%operator w
\def\hW{\hat{W}}                    %%operator W
\def\hG{\hat{G}}                    %%operator G
\def\hT{\hat{T}}                    %%operator T
\def\hnu{\hat{\nu}}                 %%operator \nu
\def\hPhi{\hat{\Phi}}               %%operator \Phi
\def\hPsi{\hat{\Psi}}               %%operator \Psi
\def\n{\noindent}
\def\bra{\left\langle}        %% big <
\def\ket{\right\rangle}       %% big >
\def\pr{\prime}              
\def\prr{\prime\prime}

\begin{document}

\title{Higher spin constraints and the super 
$\left( W_{\frac{\infty}{2}}\oplus W_{\frac{1+\infty}{2}}\right)$
algebra in the super eigenvalue model}

\author{L. O. Buffon$^{1}$, D. Dalmazi$^{2}$ and A. Zadra$^{1}$\\
{}\\
${}^{1}${\it Instituto de F\'{\i}sica, Universidade de S\~ao Paulo }\\
{\it CP 66318, 05389-970,  S\~ao Paulo, Brazil }\\
{\it lobuffon@uspif.if.usp.br, azadra@uspif.if.usp.br}\\
{}\\
${}^{2}${\it UNESP, Campus de Guaratinguet\'a }\\
{\it CP 205, Guaratinguet\'a, S\~ao Paulo, Brazil}\\
{\it dalmazi@grt000.uesp.ansp.br}
}

\date{}

\maketitle

\begin{abstract}
We show that the partition function of the super eigenvalue model satisfies 
an infinite set of constraints with even spins $s=4,6,\cdots,\infty$. 
These constraints are associated with half of the bosonic generators of the 
super $\left( W_{\frac{\infty}{2}}\oplus W_{\frac{1+\infty}{2}}\right)$ 
algebra. The simplest constraint $(s=4)$ is shown to be reducible
to the super Virasoro constraints, previously used to construct the model. 
All results hold for finite $N$.
\end{abstract}

\section{Introduction}

Some time ago, Kazakov showed \cite{1} that the discrete hermitian one-matrix
model exhibits a transition to a massless phase. In the continuum limit,
it describes the $(p,q)=(2,2k-1)$, $k=2,3,4,...$, minimal models conformally 
coupled to $2D$-gravity. One of the basic features of this model is the
presence of the Virasoro constraints satisfied by its partition function.
These constraints can be derived by various methods \cite{2,3,4}. 
Indeed, they hold even before the phase transition takes place (see for 
instance \cite{3,4}). In \cite{4}, the Virasoro constraints
were shown to be a consequence of a set of Schwinger-Dyson (S-D) equations 
associated with the differential operators 
$l_{n}=-\sum_{i=1}^{N}x_{i}^{n+1}\partial_{i}$ $(n\geq -1)$, 
where $x_{i}$ ($i=1,...,N$) are the eigenvalues of the hermitian
$N\times N$ matrix. This fact rises the following question: what is the 
r\^ole of the S-D equations associated with the higher order (or higher spin) 
differential operators $W_{n}^{s}=
\sum_{i=1}^{N}x_{i}^{n}\partial_{i}^{s-1}$ ($n\geq 0, s\geq 2$) ? This set 
contains the Virasoro generators $l_{n}$ and forms a 
$W_{1+\infty}$ algebra. As shown in \cite{5}, each
operator $W_{n}^{s}$ gives rise to a S-D equation which, on its turn,
originates a constraint on the partition function. However, such constraints 
can be reduced to the Virasoro constraints, at least for the spins $s=3,4$.

The main purpose of this work is to address the issue of higher spins 
constraints in the super eigenvalue model. This supersymmetric discrete model 
was proposed in \cite{6} as a way to describe some minimal models coupled to 
$2D$-supergravity. It is supposed to be a supersymmetric extension of the 
effective bosonic eigenvalue model. We will show that there actually is 
an infinite set of differential operators which give rise to S-D equations 
and corresponding higher spin constraints on the super eigenvalue partition 
function. However, opposing the bosonic theory, we only 
find even spin constraints ($s=4,6,8,...$). Furthermore, the corresponding 
differential operators seem to be (at least for $s=4,6$) linear
combinations of half of the bosonic generators of the super  
algebra
$\left( W_{\frac{\infty}{2}}\oplus W_{\frac{1+\infty}{2}}\right)$
\cite{7,8}. This algebra contains the $N=1$ superconformal algebra and forms 
a natural $N=1$ supersymmetric extension of the $W_{n}^{s}$ operators. The 
simplest constraint, with spin $s=4$, is worked out explicitly. 
As in the bosonic model, it can be reduced to the super Virasoro constraints.

This paper is organized as follows. In section II,  we review the results on 
the bosonic theory, based on reference \cite{5}. In section III, we obtain 
the higher spin constraints in the super eigenvalue model and prove that the
constraint $s=4$ is reducible. In section IV, we relate the super  
$\left( W_{\frac{\infty}{2}}\oplus W_{\frac{1+\infty}{2}}\right)$ algebra
to the higher spins constraints.
Section V contains a brief summary of the results and comments on 
the reducibility of the constraints with $s>4$.

\section{Higher spin constraints in the hermitian one-matrix model}

The partition function of the hermitian one-matrix model is given by:

\be
\label{A}
Z=\int {\cal D}M \exp{(-N\sum_{k=0}^{\infty}g_{k}\;Tr\;M^{k})}=
\int (\prod_{i=1}^{N}dx_{i}e^{V(x_{i})})\Delta_{N}^{2}\quad ,
\ee
where $x_{i}$ ($i=1,...,N$) are the eigenvalues of hermitian $N\times N$ 
matrix $M$; ${\cal D}M$ is the flat measure; 
$\Delta_{N}=\prod_{i<j=1}^{N}(x_{i}-x_{j})$ is the van der Monde 
determinant and
\be
V(x_{i})=-N\sum_{k=0}^{\infty}g_{k}x_{i}^{k}
\ee
is the potential, which depends on the coupling constants $g_k$. 
By making infinitesimal non-singular conformal transformations, 
$\delta x_{i}=[x_{i},\epsilon_{n}l_{n}]=
\epsilon_{n}x_{i}^{n+1}$, which are generated by the differential operators  
$l_{n}=-\sum_{i=1}^{N}x_{i}^{n+1}\partial_{i}$, one derives the Virasoro
constraints: 
\be
\label{B}
\hL_{n}Z=0\qquad ,\qquad n\geq -1\quad ,
\ee
where
\be 
\label{B1}
\hL_{n}=\sum_{k\geq 0}kg_{k}\frac{\partial}{\partial g_{n+k}} +\frac{1}{N^{2}}
\sum_{\mu=0}^{n}\frac{\partial^{2}}{\partial g_{\mu}\partial g_{n-\mu}}\quad .
\ee
The operators $\hL_{n}$ and $l_{n}$ satisfy the same algebra,  
$[\hL_{n},\hL_{m}]=(n-m)\hL_{n+m}$. 
The Virasoro constraints (\ref{B}) will be henceforth called spin two $(s=2)$
constraints. 

In \cite{5}, the authors investigated the higher spin $(s>2)$ 
constraints associated with the operators
$W_{n}^{s}=\sum_{i=1}^{N}x_{i}^{n+1}\partial_{i}^{s-1}\;\;\;(n\geq
-1\;,\;s\geq 1)$, which generate the $W_{1+\infty}$ algebra. The 
infinitesimal transformations associated with $W_{n}^{s}$ for $s>2$ cannot be 
written in a local form in $x_{i}$ configuration space. Therefore it is  
convenient to derive the constraints from the S-D equations. These equations 
follow from integrals of total derivatives, which must be suitably chosen 
because the action of the operators $W_{n}^{s}$ on $\Delta_{N}^{2}e^{\sum_{i=
1}^{N}V(x_{i})}$ is, in general, rather complicated. The solution to
this puzzle comes from the following property of the van der Monde determinant:
\be
\label{D}
\sum_{i=1}^{N}\partial_{i}^{s}\Delta_{N}=0\qquad ,
\ee
where $s\geq 1$ is an integer. 
As shown in appendix 1, this property implies\footnote{ The opposite 
is also true: from (\ref{C}) one can prove (\ref{D}) by induction.}
\be
\label{C}
\left(\sum_{i=1}^{N}\frac{1}{p-x_{i}}\partial_{i}^{s-1}\right)\Delta_{N}=
\Delta_{N}\frac{(\partial+w(p))^{s}\cdot 1}{s}\quad .
\ee
We have introduced the loop variable $w(p)=\sum_{
i=1}^{N}\frac{1}{p-x_{i}}=\sum_{i=1}^{N}\sum_{n\geq 0}\frac{x_{i}^{n}}{p^{n+1}}
$ and the notation $\partial \equiv \partial /\partial p$. 
Equation (\ref{C}) can be generalized \cite{5} as follows,
\[
W_{s}(p)\left(\Delta_{N}e^{\beta V}\right)=\Delta_{N}e^{\beta
V}\frac{[(\partial +w(p)+\beta V^{\prime}(p))^{s}\cdot 1]_{-}}{s}\quad ,
\]
\be
\label{E}
W_{s}^{\dagger}(p)\left(\Delta_{N}e^{\beta V}\right)=-(\Delta_{N}e^{\beta
V})\frac{[(\partial -(w(p)+\beta V^{\prime}(p)))^{s}\cdot 1]_{-}}{s}\quad ,
\ee
where $[f(p)]_{-}$ means only negative powers of $p$; $\beta$ is a
real constant and $W_{s}(p)$ ($W_{s}^{\dagger}(p)$) is the resolvent operator
for the differential operators $W_{n}^{s}$ ($W_{n}^{\dagger\;s}$):
\[
W_{s}(p)=\sum_{i=1}^{N}\frac{1}{p-x_{i}}\partial_{i}^{s-1}=\sum_{n\geq
0}\frac{W_{n-1}^{s}}{p^{n+1}}\quad ,
\]
\be
W_{s}^{\dagger}(p)=(-1)^{s-1}\sum_{i=1}^{N}\partial_{i}^{s-1}
\frac{1}{p-x_{i}}\quad .
\ee
The simplicity of the r.h.s. of equation (\ref{E}) suggests
\cite{5} that we take the identities
\bea
&\int& \!\!
\prod_{i=1}^{N}dx_{i}\left[\Delta_{N}e^{(1\!-\!\alpha)\sum_iV(x_{i})}
W_{s}\left(\Delta_{N}e^{\alpha\sum_iV(x_{i})}\right) \!\! -\!\!
\Delta_{N}e^{\alpha\sum_iV(x_{i})}
W_{s}^{\dagger}\left(\Delta_{N}e^{(1-\alpha)\sum_iV(x_{i})}\right)
\right]\nonumber \\
&=& 0 \qquad ,
\eea
where $\alpha$ is an arbitrary real constant. Using (\ref{E}) we have the
S-D equations for $s=2,3,4,$ respectively, 
\be
\label{F}
\bra \left[ \left( \partial \phi \right) ^2\right] _- \ket = 0\quad ,
\ee

\be
\label{F1}
\bra \left[ \left( \partial + (2\alpha -1)V'\right) \left( \partial \phi 
\right) ^2 \right] _- \ket = 0\quad ,
\ee

\bea
\label{F2}
\bra  \left[ \left( \partial \phi \right) ^4 - \left( \partial ^2 \phi 
\right)^2 + 2 \partial ^2 \left( \partial \phi \right)^ 2\right] _- \ket 
&+& 3(2\alpha -1)\bra \left[ \partial \left( V' \left( \partial \phi \right)^2
\right) \right] _-\ket \nonumber \\
&+& {3\over 2}(2\alpha -1)^2 \bra \left[ {V'}^2 \left(
\partial \phi \right) ^2 \right] _- \ket = 0 \quad .
\eea
where $V'(p)= -N\sum _k kg_k p^{k-1}$ and $\partial \phi \equiv w(p) + 
{1\over 2} V'(p)$ behaves like a spin one current. 
Equation (\ref{F}) is the so called loop equation which can be
solved perturbatively in $1/N$. Notice that, if we choose $\alpha = 1/2$, all 
equations will be written in terms of $\partial \phi $ and its derivatives 
only. 

We stress that the above equations also hold for the reduced
models, i.e. when $g_{k}=0$ for some $k>m$. However, only in the general 
case ($g_{k}\neq 0$ for any $k$), we can rewrite them as constraints on the
partition function $Z$. Using the property
\be
-\frac{1}{N}\frac{\partial}{\partial
g_{n}}e^{V}=\sum_{i=1}^{N}x_{i}^{n}e^{V}\qquad ,
\ee
and introducing the loop operator
\be
\label{2.13.b}
\hw=-\frac{1}{N}\sum_{n\geq
0}\frac{1}{p^{n+1}}\frac{\partial}{\partial g_{n}}\qquad ,
\ee
equation (\ref{F}) becomes
\be
(\hw^{2}+V^{\pr}\hw)_{-}Z= \sum_{n\geq -1}\frac{\hL_{n}}{p^{n+2}}Z
\quad ,
\ee
where $\hL_{n}$ was given in (\ref{B1}). Therefore we recover the
constraints (\ref{B}). Analogously, equations (\ref{F1}) and  (\ref{F2}) give
rise to further constraints on the partition function,
\[
\hW_{\mu}^{3}Z=0\qquad ,\qquad \mu\geq -2\qquad ;
\]
\[
\hW_{v}^{4}Z=0\qquad ,\qquad v\geq -3\qquad ;
\]
where the operators $\hW^{3}_{\mu}$ and $\hW^{4}_{v}$ can be written as
\be
\label{G1}
\hW_{\mu}^{(3)}=(\mu+2)\hL_{\mu}+ N(2\alpha-1)\sum_{k\geq
0}kg_{k}\hL_{\mu +k}\quad ,
\ee

\bea
\label{G2}
\hW_{v}^{(4)}&=&\frac{3}{2}(v+2)(v+3)\hL_{v}+ 3N(2\alpha-1)(v+3)\sum_{k\geq
0}kg_{k}\hL_{v +k}+\sum_{n=-1}^{v+1}\hL_{n}\hL_{v-n}\nonumber \\
& &+N^{2}(2+6\alpha(\alpha-1))\sum_{k,k^{\pr}}kk^{\pr}g_{k}
g_{k^{\pr}}\hL_{v+k+k^{\pr}}
- 2N\sum_{k\geq 0}\sum_{n=0}^{v+k+1}kg_{k}\frac{ 
\partial}{\partial g_{n}}\hL_{v+k-n}
\eea
As stressed in \cite{5}, the $s=3,4$ constraints are reducible
to the $s=2$ Virasoro constraints and therefore impose no further
restrictions on $Z$. It has been conjectured in \cite{5} that
this should also hold when $s>4$, although no proof is
available. Now a comment is in order: if the algebra of the $\hW_{\mu}^{(s)}$
constraints were isomorphic to the algebra of the differential operators
$W_{\mu}^{s}=\sum_{i}x_{i}^{\mu+1}\partial_{i}^{s-1}$, from which they 
indirectly come, then it should be obvious that $\hW_{\mu}^{(s)}Z=0$ for $s>4$,
because any operator $W_{n}^{(s)}$ can be obtained from 
$W_{n}^{(2)}$ and $W_{n}^{(3)}$ via commutators. However, the operators $s=2$ 
and $3$ obey the commutation relation
\be
\label{H}
[\hL_{m},\hW_{\mu}^{3}]=-2m(m+1)\hL_{\mu+m} +(2m-\mu)\hW_{\mu+m}^{(3)}
+\frac{2(2\alpha-1)}{N}\sum_{k=0}^{m}k\frac{\partial}{\partial g_{m-k}}\hL_{
\mu+k}
\ee
The first two terms on the r.h.s. of (\ref{H}) correspond to $[-\sum_{i}x_{i}
^{m+1}\partial_{i},2\sum_{i}x_{i}^{\mu+2}\partial_{i}^{2}]$, but the
last one breaks the isomorphism. These commutators may  be isomorphic
only for $\alpha=1/2$, when $\hW_{\mu}^{3} (\alpha=1/2)=(\mu+2)\hL_{\mu}$. 
However, after calculating the commutation relations between higher spin 
operators, we concluded that there is no value of $\alpha $ for which the 
algebras (of constraints and differential generators) are isomorphic.

\section{Higher spin constraints in the supereigenvalue model}

The partition function for the super eigenvalue model ($Z^S$) was introduced
in \cite{6} and reads:
\be
Z^S=\int {\cal D}\mu\;\Delta_{N}^{S}e^{\sum_{i=1}^{N}
(V(x_{i})+\psi(x_{i})\theta_{i})} \quad ,
\ee
where
\[
{\cal D}\mu=\prod_{i=1}^{N}dx_{i}d\theta_{i}\qquad , \qquad 
\Delta_{N}^{S}=\prod_{i<j=1}^{N}(x_{i}-x_{j}-\theta_{i}\theta_{j})\quad ,
\]

\be
V(x_{i})=-N\sum_{k=0}^{\infty}g_{k}x_{i}^{k}\qquad , \qquad 
\psi (x_i)=-N\sum _{k=0}^\infty \psi _k x_i^k\quad ,
\ee
If one makes infinitesimal non-singular superconformal 
transformations 
\[
\delta x_{i}=[x_{i},\epsilon_{n}g_{n+1/2}+\alpha_{n}l_{n}]=
-\epsilon_{n}\theta_{i}x_{i}^{n+1}+\alpha_{n}x_{i}^{n+1}\quad ,
\] 

\[ 
\delta \theta _i=[\theta_{i},\epsilon_{n}g_{n+1/2}+\alpha_{n}l_{n}]=
\epsilon_{n}x_{i}^{n+1}
+\frac{(n+1)}{2}\alpha_{n}\theta_{i}x_{i}^{n}\quad ,
\]
where $\epsilon_{n}$
are grassmann-odd parameters, one arrives at the following constraints:
\be
\label{H1}
\hG_{n+1/2}Z^S=0=\hL_{n}^{S}Z^S\;\;\;,\;\;\;n\geq -1\quad .
\ee
The operators $G_{n+1/2}=G_{n+1/2}(g_{k},\frac{\partial}{\partial g_{k}},
\psi_{k}, \frac{\partial}{\partial \psi_{k}})$ and $\hL_{n}^{S}=\hL_{n}^{S}
(g_{k},\frac{\partial}{\partial g_{k}},
\psi_{k}, \frac{\partial}{\partial \psi_{k}})$ can be found in the literature 
(see page 156 of \cite{9}).
They satisfy a subalgebra of the $N=1$ superconformal
algebra, which is isomorphic to the algebra of the differential
operators:
\be
\label{I1}
g_{n+1/2}=\sum_{i=1}^{N}x_{i}^{n+1}(\theta_{i}\partial_{i}- \Pi
_{i})\;\;\;\;,\;\;\;\; \Pi_{i}=\frac{\partial}{\partial \theta_{i}}\quad ,
\ee
\be
\label{I2}
l_{n}=-\sum_{i=1}^{N}\left(x_{i}^{n+1}\partial_{i}+\frac{(n+1)}{2}x_{i}^{n}
\theta_{i}\Pi_{i}\right)\;\;\;,
\ee
namely,
\[
\{g_{n+1/2},g_{m+1/2}\}=2l_{n+m+1}
\qquad , \qquad 
\{l_{n},l_{m}\}=(n-m)l_{n+m}\quad ,
\]

\be 
\{l_{n},g_{m+1/2}\}=\frac{(n-1-2m)}{2}g_{n+m+1/2}\;\;\;.
\ee
The constraints $\hG_{n+1/2}Z=0$ and $\hL_{n}^{s}Z=0$ correspond to spins
$s=3/2$ and $2$ respectively.

Inspired by the results of the previous section, it is possible 
to obtain the $s=3/2,2$ constraints from the following identity,
\be
\label{J}
\int {\cal D}\mu\;\left[e^{U}(W^{s}(p)\Delta_{N}^{S})
-\Delta_{N}^{S}(W^{\dagger\;s}(p)e^{U})\right]=0\quad ,
\ee
which is simply an integral of a total derivative. Above, 
$U=\sum_{i=1}^{N}(V(x_{i})+\psi(x_{i})\theta_{i})$;
$W^{s}(p)=\sum_{n\geq 0}\frac{1}{p^{n+1}}W_{n}^{s}$, where $W_{n}^{s}$
is the first order differential operators for $s=3/2,2$ given in  (\ref{I1})
and (\ref{I2}) respectively. These identities can be written as 
S-D equations,
\be
\label{K1}
\bra [T_{3/2}]_{-}\ket \equiv \bra [(\partial \Phi)\Psi]_{-}\ket=0\quad ,
\ee

\be
\label{K2}
\bra [T]_{-}\ket \equiv
\bra [(\partial \Phi)^{2}+ (\partial \Psi)\Psi]_{-}\ket=0\quad ,
\ee
where we have introduced the notation $\partial\Phi(p)=w(p)+V^{\pr}(p)$ 
and $\Psi(p)=\nu(p)+\psi(p)$; $w(p)=\sum_{i=1}^{N}\frac{1}{p-x_{i}}$
and $\nu(p)=\sum_{i=1}^{N}\frac{\theta_{i}}{p-x_{i}}$ are the super-loop
variables and (\ref{K1}), (\ref{K2}) are called super-loop equations
\cite{2} which give rise to the constraints (\ref{H1})
\[
:(\hat{T}_{3/2})_{-}:Z^S = [(\partial \hat{\Phi}\hat{\Psi})_{-}]Z^S
=0\quad ,
\]

\be
\label{KK2}
: [\hat{T}]_{-}:Z^S=
: [(\partial \hat{\Phi})^{2}+ (\partial \hat{\Psi})\hat{\Psi}]_{-}:Z^S=0
\quad .
\ee

\n
The operator $\hw$ was given in (\ref{2.13.b}) and
$\hnu=-\frac{1}{N}\sum_{n\geq 
0}\frac{1}{p^{n+1}}\frac{\partial}{\partial \psi_{n}}$.

Now we turn to the higher order differential operators ($s>2$). 
Since it seems that there is
no supersymmetric analog of the property (\ref{D}) for $\Delta_{N}^{S}$,
we found convenient to factorize the bosonic van der Monde determinant
$\Delta_{N}$ from $\Delta_{N}^{S}$, by writing

\be
\Delta_{N}^{S}(x_{i}-x_{j}-\theta_{i}\theta_{j})=
\Delta_{N}(x_{i}-x_{j})e^{F}\quad ,
\ee
where we define the following function:
\be
F=-\frac{1}{2}\sum_{i\neq j}\frac{\theta_{i}\theta_{j}}{x_{i}-x_{j}}\;\;\;.
\ee
For even spin $s$, there is a remarkably simple formula for the
action of some differential operator of spin $s$ (${\cal O}^{(s)}$) on
the fermionic part of $\Delta_{N}^{S}$. It is given by the equation

\be
\label{K5}
{\cal O}^{(s)}e^{F}\equiv
\sum_{i=1}^{N}\left(\frac{1}{p-x_{i}}\partial_{i}^{s-1}
+\partial_{i}^{s-1}\frac{1}{p-x_{i}} -
\partial_{p}^{s-1}\frac{\theta_{i}}{p-x_{i}}\Pi_{i}\right)e^{F}= \left[
(\partial_{p}^{s-1} \nu)\nu - \partial_{p}^{s}w\right]e^{F}
\ee
which is demonstrated in appendix 2. Higher spin constraints are thus  
obtained from the identity

\be
\label{L}
\int {\cal D}\mu
\left[e^{U}\Delta_{N}({\cal O}^{\dagger\;(s)}(p)e^{F})\right]- 
\int {\cal D}\mu
\left[{\cal O}^{(s)}(p)(e^{U}\Delta_{N})e^{F}\right]=0\quad ,
\ee
where ${\cal O}^{\dagger\;(s)}=-{\cal O}^{(s)}-(\partial_{p}^{s-1}
w)$. The second integral in (\ref{L}) can be written as a local (although 
rather complicated) function of $w$, $\nu$ and its derivatives (see appendix 
2). Finally, we obtain an infinite set of S-D equations
associated with even spin differential operators. For $s=2$, we recover
from (\ref{L}) the bosonic loop equation (\ref{K2}), which is
associated with the Virasoro constraints. For the next spin, $s=4$, we
obtain:

\be
\label{L1}
\bra \left[\frac{1}{2}(\partial \Phi)^{4} +2\partial^{3}\Phi  \partial \Phi 
+\frac{3}{2}(\partial^{2} \Phi)^{2} +\partial^{3} \Psi \Psi\right]_{-}\ket=0
\quad .
\ee
The above S-D equation can be rewritten as a constraint,

\be
\left[:\frac{\hT^{2}}{2}+ \hT_{3/2}\partial\hT_{3/2}+ \partial^{2}\hT - 
\left(\frac{(\partial^{2} \hPhi)^{2}}{2}+\hPsi^{\prr}\hPsi^{\pr}\right):
\right]_{-}Z^S=0\quad .
\ee
In order to relate it to the constraints (\ref{KK2}), we
split $\hT$ and $\hT_{3/2}$ in parts with negative and non-negative
powers of $p$, that is $\hT=\hT_{-}+\hT_{+}$, 
$\hT_{3/2}=\hT_{3/2\;(-)}+\hT_{3/2\;(+)}$. We end up with the equation

\bea
\label{LL}
& &\left[\frac{:\hT_{-}::\hT_{-}:}{2} +:\hT_{+}::\hT_{-}: +
:\hT_{3/2}^{+}::\hT_{3/2}^{\pr\;-}: - 
:\hT_{3/2}^{+\;\pr}::\hT_{3/2}^{-}: +
:\hT_{3/2}^{-}::\hT_{3/2}^{\pr\;-}: + :\partial^{2}\hT_{-}:\right.\nonumber\\
& &-\left.
:\left(\frac{(\partial^{2}\Phi)^{2}}{2}+\Psi^{\prr}\Psi^{\pr}\right): + \;\;21
\;{\rm commutators} \right]_{-}Z^S=0\quad .
\eea

We stress that $\Psi(p)$ and $\partial\Phi(p)$ behave like a two
dimensional free fermion and a spin one currents, respectively. The 
commutators between these quantities, calculated at the same point are 
ill defined in general. However, due to the projections on negative and
non-negative frequencies and because we only care for 
commutators acting on the partition function, the 
calculations can be done without ambiguities. In appendix 3 we work
out a sample calculation explicitly.

After collecting the results, we find:

\be
\label{N}
\left[ \;21\;{\rm commutators} \;-\;:\left( \frac{(\partial^{2}\Phi)^{2}}{2}
+\Psi^{\prr}\Psi^{\pr}\right) :\right] _
{-} Z^S=\left(\frac{\partial^{2}\hT}{4}\right)_{-}Z^S\quad .
\ee

\n
Therefore, from equations (\ref{LL}) and  (\ref{N}),  we conclude that the 
$s=4$ constraint on $Z^S$ is automatically satisfied as long as 
$Z^S$ already
obeys the $s=3/2,2$ constraints (super-loop equations). 
We do not have a proof of reducibility for the $s>4$ constraints.

\section{Super $\left( W_{\frac{\infty}{2}}\oplus
W_{\frac{1+\infty}{2}}\right)$
algebra and the higher spin constraints}

Instead of looking for identities like (\ref{K5}), we could have asked
ourselves what are the  differential operators that extend
the $N=1$ superconformal operators $g_{n+1/2}$ and $l_{n}$ to
infinitely higher spins, in the same way the operators
$W_{n}^{s}=\sum_{i=1}^{N}x_{i}^{n+1}\partial_{i}^{s-1}$ extend the Virasoro
generators $l_{n}$. In other words, what is the $N=1$ supersymmetric analog
of the algebra generated by $W_{n}^{s}$ $(s\geq 2)$ ? Interesting enough, 
the answer to this question seems to be unique \cite{10}. Namely, if we
define the spin of a fermionic (bosonic) differential generator as $1/2$
($1$) plus the highest power of the operator $\partial$ present in the
generator, we have the following ansatz for the generator $W_{n}^{5/2}$:

\be
\label{O}
W_{m-1}^{5/2}=\sum_{i=1}^{N}\left[x_{i}^{m}\theta_{i}\partial^{2}_{i}+
c_{m}x_{i}^{m}\partial_{i}\Pi_{i} +d_{m}x_{i}^{m-1}\Pi_{i}
+e_{m}x_{i}^{m-2}\theta_{i}\right]\quad .
\ee

\n
It is the most general ansatz \cite{10} compatible with the
spin composition rule $[W_{m}^{s},W_{n}^{s^{\pr}}]_{+}=f_{mn}^{ss^{\pr}}
W_{n+m}^{s+s^{\pr}-1} + \;\;lower\;spins$, which is obeyed by 
super $W$-algebras. If we require a closed algebra with $g_{n+1/2}$
and $l_{n}$, the coefficients in (\ref{O}) get fixed: $c_{m}=1$, $d_{m}=m$
and $e_{m}=0$. Simultaneously, a spin 2 generator ($\widetilde{W}_{m}^{2}$)
must be defined as:

\be
\widetilde{W}_{m}^{2}=
\sum_{i=1}^{N}x_{i}^{m+1}(1-\theta_{i}\Pi_{i})\partial_{i}
\quad .
\ee

No further ansatz or definitions are needed and the
higher spin generators can be obtained by the algebra 
of (anti)commutators of the generators 
$W_{m}^{s}$   $s\leq 5/2$, already defined. Curiously, 
no odd spin generators come out and there is a 
doubling of even spin generators  
at each spin level (see \cite{10} for details). For instance, at $s=4$
we have

\[
W_{n}^{4}=\sum_{i=1}^{N}\left[2x_{i}^{n+2}\partial_{i}^{3}+3(n+2)x_{i}^{n+1}
\partial_{i}^{2}+ (n+1)(n+2)x_{i}^{n}\partial_{i}-(n+2)x_{i}^{n+1}(1-\theta_{i}
\Pi_{i})\partial_{i}^{2}\right]
\]

\be
\label{P}
\widetilde{W}_{n}^{4}=\sum_{i=1}^{N}\left(1-\theta_{i}
\Pi_{i}\right)\left(x_{i}^{n+1}\partial_{i}^{3}+(n+1)x_{i}^{n}
\partial_{i}^{2}\right)\quad .
\ee

Apparently, the super algebra so obtained was first discovered by the
authors of reference \cite{7} (see also \cite{8}), where it was called
\footnote{We prefer to call it super $\left(W_{\frac{\infty}{2}}\oplus
W_{\frac{1+\infty}{2}}\right)$ algebra, since it is possible to redefine the 
bosonic generators $W_{n}^{s}$ and 
$\widetilde{W}_{n}^{s}$ such that they split in two decoupled algebras, 
$W_{\infty \over 2}$ and $W_{{1+\infty}\over 2}$, which correspond to 
truncations into even spins of the $W_\infty $ and $W_{1+\infty}$ algebras.}
Super $W_{\frac{\infty}{2}}$.
We believe that this algebra corresponds to the $N=1$ analog of the
algebra of $W_{n}^{s}$ $(s\geq 2)$ generators. We have tried, in vain,
to derive fermionic constraints on $Z^S$ from the fermionic
differential operators $W_{n}^{s}$. However, the situation for the even spin 
bosonic generators looks much better. We found
a sharp connection between the differential operators ${\cal O}^{(s)}$ of the
last section and the bosonic $W_{n}^{s}$ operators. By using
$\frac{1}{(p-x_{i})}=\sum_{n\geq 0}x_{i}^{n}/p^{n+1}$ we have, from
(\ref{K5}) and (\ref{P}),

\[
{\cal O}^{(2)}(p)e^{F}=\sum_{n\geq 0}\frac{W_{n-1}^{2}}{p^{n+1}}e^{F}\quad ,
\]

\be
\label{Q}
{\cal O}^{(4)}(p)e^{F}=\sum_{n\geq 0}\frac{1}{p^{n+1}}\left[W_{n-2}^{4}
+2n(n-1)W_{n-3}^{2}\right]e^{F}\quad ,
\ee

\[
{\cal O}^{(6)}(p)e^{F}=\sum_{n\geq 0}\frac{1}{p^{n+1}}\left[W_{n-3}^{6}
+3n(n-1)W_{n-4}^{4} +2n(n-1)(n-2)(n-3)W_{n-5}^{2}\right]e^{F}.
\]

\n
Therefore, the differential operators, which give rise to the 
higher even-spin constraints on $Z^S$,
are simple linear combinations of even spin $W_{n}^{s}$ generators
of the super  $\left( W_{\frac{\infty}{2}}\oplus
W_{\frac{1+\infty}{2}}\right)$ algebra. It must be stressed that the 
$W_{n}^{s}$ generators have very specific numerical factors in their 
definitions
(see (\ref{P})) and the above identification is a rather non-trivial result,
since ${\cal O}^{(s)}$ and $W_{n}^{s}$ were found by completely different
methods.

Some comments are in order. First, we have used the identity
$\theta_{i}\partial_{i}e^{F}=0$ to write (\ref{Q}). Therefore,
we can say that the differential operators $W_{n}^{s}$ differ from the 
differential operators 
contained in ${\cal O}^{(s)}(p)$ by terms proportional to
$\theta_{i}\partial_{i}^{m}$ $(m\geq 1)$. It would thus be impossible
to guess $W_{n}^{s}$ from ${\cal O}^{(s)}$. However, the formula (\ref{K5})
could have been obtained from $W_{n}^{s}e^{F}$. Furthermore,
since we have a doubling of even spin differential operator 
($W_{n}^{s}$ and $\widetilde W_{n}^{s}$) one might also expect a
doubling of even spins constraints at each spins $s$ level.  However, we
have not been able to get any constraint or S-D equation associated
with the $\widetilde{W}_{n}^{s}$ operators for $s> 2$. For  $s=2$ the operators
$\widetilde{W}_{n}^{2}$, which obey the same algebra of the Virasoro generators
$W_{n}^{2}$, lead to the same constraint (\ref{KK2}).

\section{Summary and final remarks}

We have derived an infinite set of even spin $(s=2r=2,4,6,...)$
constraints on the partition function of the supereigenvalue model which
include the Virasoro constraints $(s=2)$ previously found in [6]. All
constraints can be written in terms of the superloop variables $w(p)$
and $\nu(p)$ and the potentials $\psi(p)$, $V(p)$. We have 
shown that the constraints at $s=4$ are reducible to the $s=3/2$ and $s=2$
super Virasoro constraints (super-loop equations), but we do not have a proof
of the reducibility for higher spins. The situation is very similar
to the bosonic eigenvalue model. We were naturally led
to define super differential operators which are the $N=1$ supersymmetric
version of $x_{i}^{n}\partial_{i}^{s-1}$ ($s\geq 2, \;n\geq
0$). Those operators satisfy the super $\left( W_{\frac{\infty}{2}}\oplus
W_{\frac{1+\infty}{2}}\right)$ algebra, whose bosonic sector 
possesses a doubling
of even spin operators ($W_{n}^{s}$ and  $\widetilde{W}_{n}^{s}$) at each
spin $s$ level. The constraints thus obtained are associated with
linear combinations of the bosonic operators $W_{n}^{s}$. The algebraic 
meaning of such combinations is unclear.

We have two important remarks. First, all results the we have
obtained are non perturbative and hold for finite $N$. Second, although
we have not been able to derive any constraints associated with the remaining
bosonic differential operators $\widetilde{W}_{n}^{s}$ (except for 
$\widetilde{W}_{n}^{2}$ which also gives rise to the Virasoro
constraints) and the fermionic ones $W_n^s$ ($s=5/2,7/2,\cdots $),
we conjecture that such constraints do exist and are reducible. For the
same reasons we suspect that the even spin constraints for $s>4$ are also
reducible. Our conjecture is based on the super 
$\left( W_{\frac{\infty}{2}}\oplus W_{\frac{1+\infty}{2}}\right)$ algebra
and the association between the constraints $\hG_{n+1/2}Z^S=0$, 
$\hat{{\cal O}}_{n}^{s}Z^S=0$ and the differential operators 
$W_n^{3/2}\equiv g_{n+1/2}$ and $W_{n}^{s}$ ($s=2,4,6,\cdots $). 
From (anti)commutators between $g_{n+1/2}$ and 
 $W_{n}^{s}$ $(s=2,4)$ we can generate the remaining differential operators
and thus, we expect that the remaining constraints can be obtained via
(anti) commutators between $\hG_{n+1/2}$, $\hat{{\cal O}}_{n}^{s}$. 
However,  we do not expect an isomorphism between the algebras of 
constraint and differential operators. The constraints thus obtained should be
reducible to the $s=3/2,2$ super Virasoro constraints.

It must be emphasized that the complete set of couplings $(g_{k}, \psi_{k})$
is necessary to rewrite the S-D equations in the form of constraints.
Nevertheless, the S-D equations clearly hold for any finite-degree polynomial 
potential. It is therefore natural to
ask whether the reducibility of the constraints can be carried
through the corresponding S-D equation for the reduced models. We 
do not have a non-perturbative answer for this question, but at leading
order in $1/N$ it is easy to show, by using the $1/N$ factorization of
observables ($\langle AB \rangle=\langle A\rangle\langle B\rangle
+O(1/N)$) that the $s=4\;$ S-D equations are
reducible to the $s=2$ and $s=3/2$ equations. The same situation appears in 
the bosonic hermitian matrix model.

\vskip 1truecm
\noindent {\bf Acknowledgements}
\vskip 1truecm

The work of L.O.B. and part of the works of D.D. and A.Z. were supported by 
CNPq and Fapesp. 

\vskip 1truecm
\noindent {\bf Appendices}
\vskip 1truecm

\appendix
\section{1. Property (\ref{C})}

Here we demonstrate the equation (\ref{C}) starting 
from the property  (\ref{D}). We extend $\Delta_{N}$ to 
$\Delta_{N+1}=\prod_{I<J}(x_{I}-x_{J})$ by introducing an auxiliary eigenvalue 
$x_{o}=p$. The extended determinant can be written as 

\be
\Delta_{N+1}=e^{\phi}\Delta_{N}(x_{i}-x_{j})\quad ,
\ee

\n
where $\phi=\ln\;\prod_{i=1}^{N}(p-x_{i})$, and using 
$\partial_{i}^{k}e^{\phi}=0$, $k\geq 2$, eq. (\ref{D}) implies

\be
\label{R}
\partial_{p}^{s}e^{\phi}\Delta_{N}=-s\sum_{i=1}^{N}(\partial_{i}e^{\phi})
( \partial_{i}^{s-1}\Delta_{N}) - \sum_{i=1}^{N} e^{\phi}
 \partial_{i}^{m}\Delta_{N} \quad .
\ee

\n
Using once more the property (\ref{D}), for $\Delta_{N}$, and the expression  
$\partial_{i}e^{\phi}=e^{\phi}(-1/(p-x_{i}))$, equation 
(\ref{R}) finally implies

\be
\label{S}
\sum_{i=1}^{N}\frac{1}{p-x_{i}} \partial_{i}^{s-1}\Delta_{N}=\frac{\Delta_{N}}
{s}(e^{-\phi}\partial_{p} e^{\phi})^{s}=\Delta_{N}\frac{(\partial_{p}+w(p))^{
s}\cdot 1}{s}\quad .
\ee

\appendix
\section{2. On the even spin constraints}
In order to obtain the expression (\ref{K5}), we start from the equations

\be
\label{S1}
\frac{\partial}{\partial\theta_{k}}e^{F}=-e^{F}\sum_{j\neq k=1}^{N}\frac{
\theta_{j}}{x_{k}-x_{j}}\quad ,
\ee

\be
\label{S2}
\sum_{k=1}^{N}\frac{\partial_{k}^{m}}{p-x_{k}} e^{F}=
-e^{F}\sum_{j\neq k}^{N}\frac{(-1)^{m}m!\theta_{k}
\theta_{j}}{(x_{k}-x_{j})^{m+1}(p-x_{k})}\quad .
\ee
For odd $m$, the factor $\frac{\theta_{k}
\theta_{j}}{(x_{k}-x_{j})^{m+1}}$ is anti-symmetric, and we may
rewrite (\ref{S2}) as

\bea
\label{T}
& &\sum_{k=1}^{N}\frac{1}{(p-x_{k})}\partial_{k}^{m} e^{F}=
\frac{m!}{2}e^{F} \sum_{k\neq j} \frac{\theta_{k}
\theta_{j}}{(x_{k}-x_{j})^{m+1}(p-x_{k})(p-x_{j})} =\\ \nonumber
& & \frac{m!}{2}e^{F} \left[
\sum_{n=0}^{m-2}\sum_{j\neq k} \frac{(-1)^{n}\theta_{k}
\theta_{j}}{(x_{k}-x_{j})^{m-n}(p-x_{k})^{2+n}}\! +\! 
\sum_{j\neq k} \frac{\theta_{k}
\theta_{j}}{x_{kj}(p-x_{k})^{m+1}}\! -\! 
\sum_{j\neq k} \frac{\theta_{k}
\theta_{j}}{(p-x_{k})^{m+1}(p-x_{j})}\right]
\eea
where $x_{ij}= x_i-x_j$. Above, we have repeatedly used the identity:

\be
\frac{1}{p-x_{j}}=\frac{1}{p-x_{k}} -
\frac{x_{kj}}{(p-x_{j})(p-x_{k})}\quad .
\ee
The last term in (\ref{T}) corresponds to $(\partial^{m}
\nu)\nu/m!$. Using (\ref{S1}) and  (\ref{S2}) we find

\bea
\label{U}
& &\sum_{k=1}^{N}\left[\frac{2}{p-x_{k}}\partial_{k}^{m}+ \sum_{n=0}^{m-2}
\frac{m!}{(m-(n+1))!(p-x_{k})^{n+2}}\partial_{k}^{m-(n+1)}+
\frac{m!}{(p-x_{k})^{m+1}}\theta_{k}\frac{\partial}{\partial
\theta_{k}}\right]e^{F}\nonumber \\
& &=(\partial^{m}\nu)\nu e^{F}
\eea

The expression (\ref{K5}) follows immediately from (\ref{U}), for $m=s-1$.
We mention that eq. (\ref{K5}) can be further simplified (compare with 
(\ref{C}):

\be
\sum_{k=1}^{N}\left[\frac{1}{p-x_{k}}\partial_{k}^{m}+  \partial_{k}^{m}
\frac{\theta_{k}}{(p-x_{k})}\frac{\partial}{\partial\theta_{k}}\right]
e^{F}=(\partial^{m}\nu \nu)e^{F}\quad .
\ee

To derive the constraints from (\ref{L}), we also need to calculate
${\cal O}^{(s)}(e^{U}\Delta_{N})$. It is easy to derive

\be
\label{U1}
{{\cal O}^\dagger}^{(s)}(e^{U}\Delta_{N})=-({\cal O}^{(s)}+ \partial _p^{s-1}w)
(e^{U}\Delta_{N})\quad .
\ee
To calculate ${\cal O}^{(s)}(e^{U}\Delta_{N})$, 
it is sufficient to determine the expressions

\be 
\label{U2}
\widetilde{\cal L}^{(s)}(e^{U}\Delta_{N})\equiv \partial _p^{s-1}
\sum_{i=1}^{N} \frac{\theta_i}{p-x_{i}}{\partial \over \partial \theta _i}
(e^{U}\Delta_{N})\quad ,
\ee

\be
{\cal L}^{(m)}(e^{U}\Delta_{N})\equiv 
\sum_{i=1}^{N} \frac{1}{p-x_{i}}\partial _i^m
(e^{U}\Delta_{N})\quad .
\ee
As for (\ref{U2}), we notice that

\be 
\label{V}
\widetilde{\cal L}^{(s)}(e^{U}\Delta_{N})= \Delta_Ne^U\sum_{k=1}^N
\sum_{l\leq 0}
\frac{\theta_k}{p-x_{k}}\psi_l x_k^l
= - \Delta_{N}
e^{U}\partial_{p}^{s-1} (\nu(p)\psi(p))_{-}\quad .
\ee

After similar manipulations and using formula (\ref{E}), we get
from (\ref{V}) the following result:

\be
\label{V1}
{\cal
L}^{(m)}(\Delta_{N}e^{U})=\frac{\Delta_{N}e^{U}}{(m+1)}[(\partial_{p}
+w+ V^{\pr}
)^{m+1}\cdot 1]_{-} + e^{U}\sum_{l=0}^{m-1} 
\left(\begin{array}{c}
m \\
l \end{array}
\right) 
\left[G_{n-l}(V(p),\psi(p))
{\cal
L}^{(l)}_{F}\Delta\right]_{-}
\ee

\n
where

\be
G_{m-l}=\sum_{t=1}^{m-l}
\left(\begin{array}{c}
m-l \\
t \end{array}
\right) 
(\partial^{t}\psi(p)).(\partial+V^{\pr})
^{m-l-t}\cdot 1\quad ,
\ee

\n
and

\be
\label{Y}
{\cal
L}^{(l)}_{F}(p)\Delta_{N}=\left(\sum_{i}\frac{\theta_{i}}{p-x_{i}}\partial_{i}
^{l}\right)\Delta_{N}\quad .
\ee
We have not been able to write the r.h.s. of (\ref{Y}) as a function
of $w(p)$ and $\nu(p)$, but using

\be
\int {\cal D} \mu\; e^{U+F}({\cal L}^{(l)}_{F}\Delta_{N})=
\int {\cal D} \mu\;\Delta({\cal L}^{\dagger\;(l)}_{F} e^{U+F})
\ee

\n
and

\be
{\cal L}^{\dagger\;(l)}_{F} e^{U+F}=e^{U+F}\sum_{t=0}^{l}(-1)^{t} 
\left(\begin{array}{c}
l \\
t \end{array}
\right) 
\partial_{p}^{l-t}(\nu(p)(\partial+V^{\pr})^{t}\cdot 1)_{-} \quad ,
\ee
we finally obtain

\bea
\label{Z}
\int {\cal D} \mu\; e^{F}{\cal L}^{(m)}(\Delta_{N}e^{U})&=&
\int {\cal D} \mu\;
e^{U}\Delta_{N}\left[\frac{(\partial_{p}+w+V^{\pr})^{m+1}\cdot 1}
{m+1}\right. \\ \nonumber
&+& \left. 
\sum_{l=0}^{m-1}\sum_{t=0}^{l}(-1)^{t} 
\left(\begin{array}{c}
m \\
l \end{array}
\right) 
\left(\begin{array}{c}
l \\
t \end{array}
\right) 
G_{m-l}(p)
\partial_{p}^{l-t}(\nu(\partial+V^{\pr})^{t}\cdot 1)_{-}\right]_{-}
\eea

From (\ref{U1}), (\ref{V}),  (\ref{Z}) and (\ref{K5}) we can find a
closed expression for the second integral of (\ref{L}) as a function of 
$\nu(p)$, $w(p)$, $V^{\pr}(p)$, $\psi(p)$, which completes the derivation 
of the even spins constraints.

\appendix
\section{3. Commutators}

Here we present an example of the calculation of 
one of the commutators in (\ref{LL}). We will take
$\hat{C}\equiv \partial\Psi^{(+)}[(\partial\Psi^{(+)}\hPsi^{(-)})_{-},
\hPsi^{(-)}]$, 
where

\be
\Psi^{(+)}(p)=-N\sum_{k\geq 0}\psi_{k}p^{k}
\qquad , \qquad
\hPsi^{(-)}(p)=-\frac{1}{N}\sum_{n\geq
0}\frac{1}{p^{n+1}}\frac{\partial}{\partial \psi_{n}} \quad .
\ee
From such definitions, we have

\be
(\partial\Psi^{(+)}\hPsi^{(-)})_{-}=\sum_{k\geq
0} \sum_{\mu \geq -1}
\frac{k\psi_{k}}{p^{\mu+2}}\frac{\partial}{\partial 
\psi_{\mu+k}}
\quad ,\quad 
(\partial\Psi^{(+)}\hPsi^{(-)})_{+}=\sum_{k\geq
0} \sum_{\mu\geq 2} k\psi_{k} p^{\mu-2} \frac{\partial}{\partial \psi_{k-\mu}}
\quad .
\ee
Considering the action of these operators 
on the partition function $Z^S$, we obtain:

\bea
& &\frac{1}{Z^S}(\hat{C}Z^S)=\sum_{k\geq 0} \sum_{\mu\geq 2}
\sum_{\hat{\mu}\geq -1} \sum_{i=1}^{N}\frac{k(k-\mu)N\psi_{k}}{p^{\hat{\mu}-
\mu+4}}\bra x_{i}^{\hat{\mu}-\mu+k}\theta_{i}\ket\\ \nonumber
& &=
\sum_{i=1}^{N}\bra\sum_{\hat{\mu}\geq -1}\frac{x_{i}^{\hat{\mu}+1}\theta_{i}}
{p^{\hat{\mu}+2}}\sum_{k\geq 0} \sum_{\mu\geq 2}(-Nkp^{k-1}\psi_{k})
\frac{(k-\mu)x_{i}^{k-\mu-1}}{p^{k-\mu+1}}\ket=
\sum_{i=1}^{N}\bra\frac{\theta_{i}}{p-x_{i}}\left(\frac{\partial\Psi^{(+)}(p)}
{(p-x_{i})^{2}}\right)_{+}\ket
\eea
Now, from the properties

\be
(f(p))_{+}=\frac{1}{2\pi i}\oint_{w>p}\frac{dw}{w-p}f(w)\quad ,\quad 
(f(p))_{-}=-\frac{1}{2\pi i}\oint_{w<p}\frac{dw}{w-p}f(w)\quad ,
\ee
we find

\bea
\hat{C}Z^S&=&Z^S\bra \frac{\partial^{2}\Psi^{(-)}\Psi^{(+)}}{2}-
\frac{\partial^{2}}{2}(\Psi^{(-)}\partial \Psi^{(+)})_{-} -
\partial[\Psi^{(-)}\partial^{2} \Psi^{(+)}]_{-}\ket
\nonumber \\
&=&\frac{Z^S}{2}\bra (\partial^{2}\Psi^{(-)}\partial
\Psi^{(+)})_{+} +
(\Psi^{(-)}\partial^{3} \Psi^{(+)})_{-}\ket \quad .
\eea

Therefore, while acting on the partition function $Z^S$, the operator 
$\hat C$ reads

\be
\hat{C}= -\frac{1}{2}(\partial \Psi^{(+)}\partial^{2}
\hPsi^{(-)})_{+} -
\frac{1}{2}(\partial^{3} \Psi^{(+)} \hPsi^{(-)})_{-}\;\;\;.
\ee

\end{document}